%% file: main.tex
\documentclass[sigconf]{acmart}

\AtBeginDocument{%
  \providecommand\BibTeX{{%
    \normalfont B\kern-0.5em{\scshape i\kern-0.25em b}\kern-0.8em\TeX}}}

\setcopyright{acmcopyright}
\copyrightyear{2018}
\acmYear{2018}
\acmDOI{XXXXXXX.XXXXXXX}

\acmConference[Conference acronym 'XX]{Make sure to enter the correct
  conference title from your rights confirmation emai}{June 03--05,
  2018}{Woodstock, NY}
\acmBooktitle{Woodstock '18: ACM Symposium on Neural Gaze Detection,
 June 03--05, 2018, Woodstock, NY} 
\acmPrice{15.00}
\acmISBN{978-1-4503-XXXX-X/18/06}

\usepackage[nolist,nohyperlinks]{acronym}
        \acrodef{api}[API]{Application Programming Interface}
        \acrodef{uid}[UID]{User Identifier}
        \acrodef{apk}[APK]{Android Application Pack}
        \acrodef{http}[HTTP]{Hypertext Transfer Protocol}
        \acrodef{json}[JSON]{JavaScript Object Notation}
        \acrodef{pii}[PII]{Personally Identifiable Information}
        \acrodef{os}[OS]{Operating System}
        \acrodef{osn}[OSN]{Online Social Networks}
        \acrodef{ide}[IDE]{Integrated Development Environment}
        \acrodef{url}[URL]{Uniform Resource Locator}
        \acrodef{gps}[GPS]{Global Positioning System}
        \acrodef{xml}[XML]{Extensible Markup Language}
        \acrodef{nlp}[NLP]{Natural Language Processing}
        \acrodef{svm}[SVM]{Support Vector Machine}
        \acrodef{iot}[IOT]{Internet of Things}
        \acrodef{ml}[ML]{Machine Learning}

\begin{document}

\title{Assessing Mobile Application Privacy: A Quantitative Framework for Privacy Measurement}

\author{João Marono}
\email{Jmarono@ua.pt}
\affiliation{%
  \institution{University of Aveiro, DETI}
  \country{Portugal}
}

\author{Catarina Silva}
\email{c.alexandracorreia@ua.pt}
\orcid{0000-0002-7969-8813}
\affiliation{%
  \institution{University of Aveiro, DETI}
  \country{Portugal}
}

\author{João P. Barraca}
\email{jpbarraca@ua.pt}
\orcid{0000-0002-5029-6191}
\affiliation{%
  \institution{Instituto de Telecomunicações \\ University of Aveiro, DETI}
  \country{Portugal}
}

\author{Vitor Cunha}
\email{vitorcunha@ua.pt}
\orcid{}
\affiliation{%
  \institution{Instituto de Telecomunicações \\ University of Aveiro, DETI}
  \country{Portugal}
}

\author{Paulo Salvador}
\email{salvador@ua.pt}
\orcid{}
\affiliation{%
  \institution{University of Aveiro, DETI}
  \country{Portugal}
}

\newcommand{\q}[1]{``#1''}

\begin{abstract}
The proliferation of mobile applications and the subsequent sharing of personal data with service and application providers have given rise to substantial privacy concerns. Application marketplaces have introduced mechanisms to conform to regulations and provide individuals with control over their data. However, a notable absence persists regarding clear indications, labels or scores elucidating the privacy implications of these applications. In response to this challenge, this paper introduces a privacy quantification framework. The purpose of this framework is to systematically evaluate the level of privacy risk when using particular Android applications. The main goal is to provide individuals with qualitative labels to make informed decisions about their privacy. This work aims to contribute to a digital environment that prioritizes privacy, promotes informed decision-making, and endorses the privacy-preserving design principles incorporation.
\end{abstract}

\ccsdesc[500]{Do Not Use This Code~Generate the Correct Terms for Your Paper}
\ccsdesc[300]{Do Not Use This Code~Generate the Correct Terms for Your Paper}
\ccsdesc{Do Not Use This Code~Generate the Correct Terms for Your Paper}
\ccsdesc[100]{Do Not Use This Code~Generate the Correct Terms for Your Paper}

\keywords{Privacy Quantification, Mobile Application, Android, Risk Management, Privacy Score.}


\received{20 February 2007}
\received[revised]{12 March 2009}
\received[accepted]{5 June 2009}

\settopmatter{printfolios=true}

\maketitle

\input{sections/section1.tex}
\input{sections/section2.tex}

\input{sections/section3.tex}

\input{sections/section4.tex}
\input{sections/section5.tex}
\input{sections/section6.tex}

\begin{acks}
This project has received funding from the European Union’s Horizon Europe research and innovation programme under grant agreement No 101095933 (RIGOUROUS project).
\end{acks}

\bibliographystyle{ACM-Reference-Format}
\bibliography{refs}

\end{document}

%% file: sections/section1.tex
\section{Introduction}
\label{sec:introduction}


The proliferation of mobile applications, has resulted in a high volume of data collection and sharing, amplifying privacy concerns among users. Although the permission system used in mobile devices allows users to control data access, it possesses inherent limitations. A ''privacy paradox`` prevails, wherein users often prioritize other factors than privacy when selecting applications. Furthermore, numerous popular applications lack comprehensive privacy policies.

The absence of transparency concerning data practices in mobile applications exposes users to potential privacy risks. Addressing these challenges underscores the critical need for a system that assesses and describes the Android application privacy practices. In this regard, it empowers users to make well-informed decisions.

Such a system empowers users to align their application choices with their privacy preferences and pushes application developers to support robust data handling standards. Despite existing safeguards, such as permissions and privacy policies, the absence of clear privacy metrics leaves users exposed, particularly those lacking technical expertise. 

This is connected with an overall strategy of designing privacy and security controls that are increasingly human-centric. Starting from the mobile applications, and then moving to the applications operating in edge systems (such as in 5G and 6G), other devices (cars, devices), and then the core systems, privacy should be a design factor, and transparent secure data processing must prevail. Users, as the owners of much of the data produced, should be aware and retain control of their data, which we are pursuing. 


In this paper, we intend to provide users with enhanced control, foster developer accountability, and introduce transparency in a digital landscape that values privacy, starting with mobile applications. With this aims, we propose a privacy quantification framework to evaluate the privacy risk of mobile applications.

The primary goal of the proposed approach is to establish a labeling system that employs an application's source code and permissions to educate users about privacy considerations while using Android applications. Moreover, this system aims to identify potential privacy risks within complex applications, even when they employ obfuscation or other strategies to obscure their behavior. The work can be applied to any platform, but for the sake of practicality, it is focused on the Android ecosystem.

The main contributions of this paper are:

\begin{itemize}
    \item Raise the privacy awareness of users by promoting the creation of software that respects their privacy and provides them with more control over the software they use;
    \item Develop \ac{api} permission-methods-pii mappings from the latest versions of Android, which can be used to assist others in dealing with permissions and \ac{api} methods related to privacy;
    \item Propose an Android application that can analyze other applications, and then provide privacy scores to the users;
    \item Specify a formula that can be used to calculate the privacy score of an Android application.
\end{itemize}

We advocate open-source and reproducible results. The code is available on GitHub~\footnote{\url{https://github.com/ATNoG/AndroidPrivacyQuantification}}.

The remaining paper is organized as follow: \autoref{sec:background} provides an explanation of the background and prior research; \autoref{sec:proposed_approach} clarifies the proposed approach; \autoref{sec:implementation} and \autoref{sec:results} describe the implementation details and the obtained results, respectively; and \autoref{sec:conclusion} encapsulates the primary conclusions and ensuing discussion.

%% file: sections/section2.tex
\section{Background}
\label{sec:background}

\label{current_android_privacy_landscape}

Android plays a significant role in the evolving landscape of mobile device privacy, influencing how consumers interact with their smartphones and tablets. Its open-source nature has fostered innovation, enabling developers to create diverse applications that address consumer needs. In 2019, Android was the most widely used mobile operating system, with 2B active users in their store~\cite{privacy_issues_of_android_application}. This diversity, however, poses challenges in maintaining uniform security and privacy standards across the Android ecosystem. Google has proactively addressed these issues by introducing privacy-centric features, enhancing permission controls, and maintaining security through regular updates. The goal is to create a balance between user privacy and developer flexibility for a seamless and secure mobile experience~\cite{privacy_issues_of_android_application}.

Android employs a security control system based on the Linux user separation mechanism to protect applications from external threats and unauthorized access to user data. This mechanism assigns a unique \ac{uid} to each user, determining their access to system resources. When a user installs an application on a device, it receives a UID distinct from other apps already on the system~\cite{uid_security_model}.

Android's \ac{os} utilizes a permission system to restrict each app's access level while running on a mobile device. Permissions are specified in the \textit{AndroidManifest.xml} configuration file and fall into four categories: normal, dangerous, signature, and privileged. Normal permissions are granted without user consent, while dangerous permissions prompt users for approval during app installation or at runtime~\cite{Security_centric_ranking_algorithm_two_privacy_scores_to_mitigate}.

Despite recent improvements, the permission system's challenges persist, notably the lack of granularity and unclear scope for permissions. For instance, the location permission provides broad access to location data, but the user is not informed about the specific data accessed~\cite{privacy_permissions_ios_android}.

Android relies on an extensive \ac{api} framework that acts as a security layer between applications and the underlying system resources. Apps request access to specific resources through \ac{api} methods, and users grant permission for these methods. The \ac{api} methods are documented by Google, clarifying their functions, data retrieval, and required permissions, establishing a direct link between user-granted permissions and the app's source code~\cite{framework_detecting_privacy_policy}. In addition, many apps incorporate third-party libraries to streamline development. While this practice simplifies feature implementation, it introduces privacy risks for users~\cite{libradar}.

To address privacy concerns, Google Play Store has introduced data safety and privacy policy requirements for all new applications and updates. Data safety provides users with insights into data practices such as collection, usage, sharing, and protection. Privacy policies inform users about how an app collects, stores, and discloses their data, often making data practices legally acceptable when explicitly stated~\cite{policylint}. However, a significant percentage of users remain indifferent to their privacy due to the complex nature of these mechanisms. Researchers employ various strategies of analysis to identify problems and develop solutions to enhance privacy mechanisms. These analyses serve as the initial step in understanding and addressing privacy-related challenges.



\section{Related Work}
\label{subsec:related_work}
As we aim to score applications, one important aspect is being able to analyse said applications, going beyond the application descriptions, and actually looking at the application code and behaviour. 

Analyzing Android applications involves two primary approaches: static analysis and dynamic analysis. Static analysis involves examining the application's source code, resources, and assets without running it. This method may include analyzing permissions~\cite{understanding_the_purpose_of_permission}, \ac{api}s~\cite{framework_detecting_privacy_policy}, and privacy policies~\cite{privot} to understand the app's behavior and data handling. 

In contrast, dynamic analysis evaluates the application's behavior while running. This approach involves observing how the app behaves during execution, including a network analysis~\cite{meddle, mobipurpose} and dynamic taint analysis~\cite{taintdroid, copperdroid}. The dynamic analysis provides insights into how the application functions in real-time.

\subsubsection{Permissions Analysis}

Android apps can acquire access to system resources such as the camera, \ac{gps}, Bluetooth, phone functions, network connections, and other sensors by using the permission mechanism. Users offer such permissions to apps when they install it and in recent versions during run-time. Despite Androids' permission system and strict security control, data leakage and misuse are still conceivable due to the low granularity of the permissions and the ambiguity of the sentences displayed to consumers when they need to accept a permission~\cite{framework_detecting_privacy_policy, permissions_warning_behavior}. Many solutions have focused on analyzing app permissions to improve the permission system.

\ac{nlp} approaches have been proposed to infer permission use from app descriptions such as WHYPER~\cite{whyper}, AutoCog~\cite{autocog}, ASPG~\cite{aspg}, ACODE~\cite{acode}. To detect which sentences in the description imply the use of permissions, they construct a permission semantic model. These works can find conflicts between the description and the requested permissions by comparing the output to the requested permissions. However, the data indicate that it is impossible to comprehend why permissions are utilized in more than 90\% of apps based on app descriptions. Similarly, SecuRank~\cite{securank} gives alternative apps based on the app's description of already installed apps extracting the functionalities in the app's description and the permissions to identify groups of apps with similar functionality. This solution has the same issue as the others in that it relies on the description to define the app's features. That is difficult because most app descriptions are only one paragraph long, and developers usually are not technical in them, not providing enough information. 

Other approaches focus on increase the users privacy awareness. Vallee et al.~\cite{per_mission} developed two apps to aid users when installing applications and with already installed applications based on permissions. The first app is a privacy-aware app store that assists users in finding more privacy-respecting apps. For each app, they provide its privacy rate based on the permissions of the app and describe each permission the app has. They then take the user to its correspondent Google Play Store page to install it. The other app helps users manage their already installed applications based on the app's permissions. Moreover, Struse et al.~\cite {permissionwatcher} developed a mobile app to provide users with information about other apps based on the app's permissions and set of rules to evaluate the risk. Zhu et al. ~\cite{mobile_app_recommendations} has developed a similar mobile application that recommends apps based on the app's requested permission. It provides information about the apps, such as popularity, app security risks, user preferences, and permissions and their purpose. Finally, Wijesekera et al.~\cite{Wijesekera} took it to another level by creating a classifier that makes decisions on behalf of the user based on the user's past decisions. 
The authors reated a system that allows reasonable resource requests without additional user intervention, avoiding inappropriate resource requests, prompting the user only when the system is unsure.

Even though these works tried to solve the permissions abuse of some apps, they cannot realize just through the permissions an app uses what kind of information is transmitted and used within the app. Furthermore, the most they can do is inform users about the impact permissions can have and give examples of apps in the same category that do not require so many permissions. Still, the recommendation system fails because not all apps have the same functionalities, and it is impossible to get an idea of all functionalities through the description. Another issue these works fail to address is the low granularity of the apps. An application with \verb|android.permission.INTERNET| permission can either communicate with a legitimate \ac{api} or have bad intentions and exfiltrate sensitive information. All this with the simple acceptance of a single permission. In order to guarantee that the app only accesses the necessary information, we need to examine its \ac{api} calls. This enables us to confirm that the requested permissions do not compromise the user's privacy.

Some works have integrated \ac{api} analysis into permission analysis for more accurate results. Permissions are connected to methods because the purpose of a permission is to allow an app to use \ac{api} methods to get system resources. To select the minimum number of permissions, Vidas et al.~\cite{vidas} developed an Eclipse \ac{ide} plugin called Permission Check tool that analyses the \ac{api} references in an app and cross-checks those references to a database that provides the minimum set of permissions. The tool helps developers minimize privacy issues and create a privacy-respecting app. Similarly, Bello-Ogunu et al.~\cite{permitme} created an Eclipse \ac{ide} plugin called PERMITME, which uses the method-permission mappings from other works~\cite{pscount, stowaway} to maintain the concept of least privilege by providing the minimum permissions list. The authors tested the tool with students and found that it reduced the time spent assessing the permissions.

Using the app's description and Latent Dirichlet Allocation (LDA), CHABADA~\cite{chabada} collects the main topics of the apps(Weather App, for example) and then clusters the apps in the same topic. Once they have all the apps in clusters, they extract only the sensitive \ac{api}s, meaning methods that require the user to grant permission. To obtain the set of sensitive \ac{api}s, the authors use the method-permission mappings from another work~\cite{stowaway}. Finally, they use unsupervised One-Class \ac{svm} anomaly classification to identify outliers that use \ac{api}s that are not common for that cluster. 

Felt et al.~\cite{stowaway} developed a tool called Stowaway that analyzes the application's use of \ac{api} calls, Intents, and Content Providers and uses a permission map, that identifies the permissions required for each method in the Android \ac{api}, to determine what permissions those operations require. The goal of this tool is to detect overprivileged in Android applications. The mappings used in this work were used in other works that used permissions and \ac{api} analysis. Johnson et al.~\cite{johnson} developed an architecture that analyses an app and judges if the requested permissions match the \ac{api} methods. They analyzed 141,372 apps and found that developers over-specify and under-specify the permissions in their apps.

Using both types of analysis, we can determine what sensitive information an application requests. However, the research focuses solely on identifying apps with the least set of permissions or finding over-privileged apps for their category. Unfortunately, none of the research focuses on identifying the \ac{pii}s that these applications are collecting. As a result, the research identified a significant gap in the analysis of application privacy.

\subsubsection{Third-party Libraries Analysis}

One popular feature that Android applications have is the use of third-party libraries. Third-party libraries are software components that can be included in the Android project. 
They account for a large portion of the code, having a tremendous impact on the analysis of the application~\cite{libradar}. They represent noise for some works and need to be removed to keep the integrity of the analysis~\cite{app_clone_1, app_clone_2}. Finding third-party libraries from compiled binary code with accuracy is a challenge. Several works used a whitelist of packages to identify third-party libraries~\cite{centroid, adrisk, addroid, adrob, ads_detection_1, Stevens2012InvestigatingUP, ads_detection_2, book_ad_libs}. 
Since the whitelist approach only compares the package names, it is a faster method to identify libraries. However, since obfuscation is popular nowadays, it is not possible to identify libraries just for the name of the directories because they are obfuscated. 

To surpass this limitation, AdDetect~\cite{addetect} and PEDAL~\cite{pedal} use a machine learning classifier to detect advertising libraries with a high accuracy rate. However, they do not account for other types of libraries, making it unfeasible to use for a common app that commonly uses more than ten libraries~\cite{libradar}. Libraries are crucial in the detection of app cloning. Since the same library code can be in multiple apps, app clone detection, works tend to exclude them in order to analyze only the app code. AnDarwin~\cite{andarwin} and Wukong~\cite{wukong} are both app clone detection works that use clustering to find similar apps, and both have a feature to detect third-party libraries. The limitation of these works is that they need a large set of applications to build a repository of libraries, and the detection is not as fast as the whitelist approach. 

Finally, there is a work that aims to have the accuracy of the cluster approach and the speed of a whitelist approach. Libradar~\cite{libradar} is a third-party library detection tool for Android apps that preprocessed 1 million apps and extracts features used to compare instantly against apps. Libradar can detect libraries even if their package names are obfuscated. Since third-party libraries can use \ac{api} methods directly, their code should not be avoided in our work and is important to be analyzed as much as the main app source code.

\subsubsection{Privacy Policy Analysis}

Privacy policies provide legal information about how data is data collected, shared, and stored by the app. If every app was transparent, using such a document should be enough to know the data practices of all the applications in the marketplace. However, such a statement is not valid. Due to this, several works emerged to analyze the current landscape of privacy policies in the Android ecosystem. One fundamental problem with the privacy policies is their readability. They are usually created by legal experts, which makes the statements complex for an average user to understand them fully.

To help users understand data practices, some autonomous approaches focus on evaluate privacy policies by using \ac{ml} and \ac{nlp} strategies, such as PrivacyGuide~\cite{privacyguide} that summarizes privacy policies; PrivOnto~\cite{privonto} represent annotated privacy policies; or ~\cite{catarina_privacy_policy_analysis} that evaluate the privacy policy compliance level in eHealth applications.  

Privacy policy analysis should be enough to understand the mobile app's data practices. However, this document structure does not follow a standard model and can have some limitations, like contradictions and ambiguous statements.
These contradictions can significantly impact the final result of the analysis since they can lead to incorrect interpretation of data practices. 
Prior approaches started to tackle this issue, however, they fail to account for all the cases of not detecting complex statements. Some examples used crowdsourced ontologies~\cite{contradictions_1, contradictions_2} and others rely on keyword-based techniques that use bi-grams and verb modifiers to detect negative statements~\cite{contradictions_3,contradictions_4}. To fill the gaps in these works, policylint~\cite{policylint} used \ac{nlp} to identify contradictions in different semantic levels of granularity. From 11,430 Google Play apps, they identified logical contradictions within 17\% of privacy policies. 
To solve the issue of non-standard privacy policies, Rowen et al.~\cite{page} created an Eclipse \ac{ide} plugin that generates a privacy policy by asking questions about the app's implementation. It then uses a privacy policy template alongside the answers. Similarly,~\cite{cranor2002platform} have attempted to bridge the gap by standardizing privacy policies but failed to reach privacy policies for mobile apps.


Some works~\cite{privot} have combined privacy policy analysis with network analysis. PriVot~\cite{privot} considers two main components: a privacy policy automatic analyzer that creates a summary of the app's privacy policy and a network analysis tool that analyzes the app network traffic. This tool notifies the user if sensitive information is captured in the packets. Other works~\cite{framework_detecting_privacy_policy} combined the privacy policy analysis with \ac{api} analysis. Salvin et al.~\cite{framework_detecting_privacy_policy} created mappings of privacy policy statements to \ac{api} methods based on real-world app privacy policies and \ac{api} documentation. They proposed an approach that validates the privacy policy statements against the source code based on an ontology of about 368 phrases. The approach proposed by the authors has similar limitations to the ones presented in prior works related to privacy policy analysis. One of the limitations is that they cannot distinguish closely related information such as \textit{device id} and \textit{device information}.

\subsection{Privacy Score}

A privacy score is a measure or rating that quantifies a product, service, application, or platform's level of privacy risk. It attempts to assist users in making educated decisions about their online interactions by measuring the privacy of their personal information. 
Existing works on creating a privacy score are few and new, indicating that this subject is relatively underdeveloped. The first works exploring this field came from Liu and Terzi~\cite{privacy_score_1, privacy_score_2}. They developed a privacy score for the potential risk a user can have by participating in \ac{osn}. They based their formula primarily on two variables: sensitivity and visibility. The score satisfies the following two properties: The more sensitive information a user provides, the greater the risk to his privacy. Also, the more visible the information, the higher the risk. Researchers believe that if more people are willing to disclose some information, then it becomes less sensitive. Due to that, they developed an equation expressing how willing a user is to disclose some information. However, they account that every user sees the sensitivity of each item as the same. Their formula is somewhat limited because they only take into account the profile's piece of information sensibility and visibility. Srivastava and Geethakumari~\cite{privacy_score_3} developed an extension for the previous formula that also examined messages to extract information and then quantify how much that information was exposed to the user. Another study examining the privacy score in \ac{osn} comes from Nepali et al.~\cite{privacy_score_4, privacy_score_5}. They developed the \verb|Privacy Index|, which is a formula that describes that are items published and not published. They calculate the sensitivity of the published items and divide by the sensitivity of all items~(published or not). PScore~\cite{pscore} takes another approach, assuming that privacy concerns differ for each user and considers the information implicitly available.

Previous work is a reasonable basis for developing the formula that calculates the privacy score of applications. However, they are related to risk assessment in \ac{osn} 's, which deals with another type of environment and associated information. Types of metrics, such as seals or privacy scores that represent the mobile app's privacy, can dramatically minimize the perceived risk~\cite{A_Longitudinal_Study_of_Information_Privacy_on_Mobile_Devices}. Due to that, some works started to develop formulas to calculate the privacy score of mobile apps to bring users more awareness and transparency. Mohsen et al.~\cite{Security_centric_ranking_algorithm_two_privacy_scores_to_mitigate} developed an intrusiveness score using the app permissions, receivers, and users privacy preferences. Even though the authors considered the user's preference and the app details, their static metrics are based only on app permissions and receivers, which are not enough indicators to compute an accurate privacy score. By only looking at this information, it is impossible to know what the \ac{pii}s that the applications are requesting to the device.
Similarly, Gates et al.~\cite{gates} used app permissions and machine learning to develop a risk score function. The authors describe that previous approaches tried to define applications as risky or not risky, which proved relatively limited. Instead, they define a numerical score based on the app permissions and details.

Other works used the Androrisk module \footnote{\url{https://github.com/vivainio/androguard/blob/master/androrisk.py}} from a popular tool called Androguard~\cite{androrisk, androrisk_2, androrisk_3}
This module determines an app's security and privacy risk level based on dangerous permissions and the presence of dangerous functionalities used by the application. It uses fuzz logic
and 21 risk categories to compute a risk score between 0 and 100. Androrisk defines weights for each permission \footnote{\url{https://github.com/androguard/androguard/blob/008f1dc438fb0a4cb846045fbbbb180115ad9224/androguard/core/analysis/risk.py}} but does not account for the potential \ac{pii}s handled by the application not translating to an accurate privacy score. 

Most works that have developed a privacy score only consider an Android application's permissions. This indicator should not be used alone because each permission has a broad set of capabilities, making it inaccurate. In addition to the permissions, the \ac{api} methods that extract information from the user's device should be known, as well as what information is associated with those methods and how dangerous that information might be to the user. Only once this information is available can a more accurate score be developed for the privacy that the application provides to the user.

%% file: sections/section3.tex
\section{Privacy Quantification of Mobile Applications}

The research on mobile privacy risk can be split into three categories: Mobile Application Permission Analysis, Mobile Application Privacy Policy Analysis, and Mobile Security and Privacy Framework. Mobile Application Permission Analysis involves examining the Android Manifest file to determine whether permissions pose a threat to user privacy. Mobile Application Privacy Policy Analysis involves analyzing an application's mandatory and public privacy policy. The aim is to identify inconsistencies between what the app gathers and what is stated in the privacy policy. The final category focuses on various types of analysis, including static analysis of the app source code and network analysis of the app requests.

An automatic privacy solution was necessary due to the nature of the problem. Thus, only static analysis was feasible at this stage. Between the different approaches in the static analysis, we chose the Application Permission Analysis and the Source Code Analysis. Using this two types of analysis, we can get an idea of what methods and permissions the application is requesting and the information associated with them.

Source Code Analysis is limited by obfuscation, which is a software development method that makes the packaged binary code more difficult to comprehend and analyze. It involves modifying the code in a way that keeps it functional but makes it difficult for people to understand. Some of the techniques include renaming methods and variables. However, it is not recommended to obfuscate the \ac{api} classes and their methods because they can cause problems and make debugging more difficult for developers.

\ac{api} methods are used by apps to communicate with the underlying system and request resources like camera and location. They are crucial in terms of privacy since they have access to the primary services that can identify the user. Since these methods are usually not obfuscated and can create a picture of what information the app collects from the device, they are a good resource to use in order to calculate the privacy of an app. To supplement the information extracted from the methods, we extract the permissions presented in the Android Manifest file.

Our approach involves scrutinizing the entire application for pertinent public methods and permissions and utilizing them to gauge privacy. To perform a search in the application, our solution uses pre-existing datasets. In order to locate \ac{api} methods and permissions related to privacy, it was imperative to identify the classes and their corresponding methods that posed a potential threat to the user's personally identifiable information, along with hazardous Android permissions that required consideration. We segregated this into two separate datasets: one encompassing the \ac{api} classes' public methods and the other containing the Android manifest's permissions.

To begin with the initial dataset, we delved into a study that scrutinized 1090 apps across 13 categories on the Google Play Store, with a focus on public methods concerning privacy~\cite{state_of_the_art_2}. We started by utilizing the inventory of methods and classes employed by that study~\cite{state_of_the_art_2}. However, we soon realized that specific methods were obsolete, while others failed to gather privacy information. Furthermore, there were new methods related to privacy that were pertinent to our study. Since every method has a purpose described in the documentation, we begin by looking at the method's description of each class and assign a \ac{pii} based on the personal information that can be retrieved using each method.

Following the development of the dataset using the public methods and \ac{pii}s related to them, the subsequent step involves devising a plan to utilize the gathered data to create a score that gauges the application's privacy based on these methods and the permissions. For each level of information sensitivity, we used a previous data sensitivity categorization~\cite{score_catarina_study} that divides data into five privacy levels: Sensitive, Personal, Confidential, Public, and Non-personal. Sensitive data can harm the user if it is disclosed, Personal data is specific to a person and requires safeguarding, Confidential data might not be personal but could be sensitive and needs safeguarding, Public data is accessible to the public and does not need safeguarding, and Non-personal data is not associated with a person and does not present any privacy concerns. 

To calculate the final privacy score, we also assigned weights to each privacy level based on~\cite{score_catarina_study}: 40 for the sensitive level, 30 for the personal level, 15 for the confidential level, 10 for the public level, and 5 for the non-personal level. Fewer transitions occur between non-personal and public levels than between confidential, personal, and sensitive levels because handling more sensitive data requires greater detail. A more detailed analysis in future work can improve the values mentioned.

After creating the essential datasets to aid in the scanning process, the first step is to extract the files from the \ac{apk} downloaded by the Application Download component. We use a tool called JADX \footnote{\url{https://github.com/skylot/jadx}}. JADX is a decompiler designed for Android apps, which enables developers to reverse-engineer compiled Android applications~\ac{apk} and retrieve the original Java source code. JADX extracts the files to two directories: resources and sources. The resources directory contains various assets used by the Android application. Some of these assets are icons, images, animation files, and the Android manifest. The sources directory has the decompiled Java source code of the application, and resource-related Java files. We search for the public methods in the sources directory and permissions in the resources directory.

Once the source code of the \ac{apk} has been decompiled into the sources directory, we start analyzing the application. The analysis process consists of recursively iterating over every directory in the sources directory, looking for valid files. For each file found, we read its contents, and if it has the class name in the import section, it means that it is using some public method of that class. We iterate over each class in our public methods dataset, and if one is found, we search for each public method of that class. If a method related to privacy is found, we verify if we have already encountered that method in the analysis of this application. We do this because if we have already found the same method, that means the application is already collecting the information retrieved from that method. If the method was not previously processed, we add it to a list of methods related to privacy, with the \ac{pii}s and privacy level related to them. At the end of this step, we have a list of public methods used by the app that are collecting private information.

For the app permissions, we do a similar process, but instead of iterating between directories, we collect all the permissions in the \textit{AndroidManifest.xml} file. The \textit{AndroidManifest.xml} is a \ac{xml} document that encompasses various components such as Activities, Permissions, Intent Filters, and Application Metadata. The element of this \ac{xml} file we want to extract is named \textit{uses-permission}. The \textit{uses-permission} element is utilized to announce the permissions necessary for an Android app to reach specific device resources. It contains a crucial attribute called \textit{android:name}, which specifies the name of the required permission. This attribute corresponds to a system permission, so the value passed in it represents the name of the permission that the app needs to request from the user in order to function properly.

After identifying the permissions required by an Android app, the first step is to validate which permissions were used by the public methods. In the Android manifest file, each \ac{api} public method that needs access to a secure resource must specify the relevant permission for that resource. Therefore, we can eliminate the permissions that were used by those methods to access the protected resources as we know the purpose for which those permissions were utilized. The next step involves ensuring that the remaining permissions do not pose any privacy concerns. This process involves cross-checking the permissions dataset for entries corresponding to the names extracted from the application's manifest file. If there is a match, the relevant details, such as the attribute name, associated \ac{pii}, and privacy level, are stored in a list.

At the final stage of our Privacy Quantification module, we have the Score Generator.
Our score creation process includes three steps. The first step involves collecting the privacy levels of the permissions \ac{pii}s and calculate the permissions score. We combine all the privacy levels of the app's permissions and divide it by the maximum permissions an app can have in terms of privacy. Then we multiply the result by 100 and round it to have a permissions score between 0 and 100. 

We repeat the same process for the public methods collected. To calculate the method score between 0 and 100, we add up the privacy levels of all the app's methods and divide by the maximum public methods an app can have. Then, we multiply by 100 and round off the final result.

To calculate the final score of an application, we combine two values: the app permissions score and the app methods score that we previously generated. First, we sum the permissions score and methods score, which gives us a value between 0 and 200. Then, we divide this value by 200, which is the maximum score an app can have. This quotient is multiplied by 100 to get a number between 0 and 100. Finally, we round this number to the nearest integer to get the final score. To reverse the final number, meaning that the higher the number, the more private the app is, we subtract the integer we obtained from 100.


%% file: sections/section4.tex


\section{Architecture of proposed solution}
\label{sec:proposed_approach}

In our work we present an approach that identifies the types of data exposed by Android mobile applications and quantifies the privacy of these applications based on the data obtained through static analysis. The result is a privacy score to assess the level of privacy of the application.

Our methodology does not focus on a specific scenario but covers a wide range of applications that require privacy analysis. Therefore, it relies on an automated tool that uses static analysis to eliminate the need for user interaction. 

We focus on the Android api classes as they are used to communicate with the underlying OS, and access services such as the camera and microphone assuming a static analysis. We also analyze the Android permissions of the application specified in the manifest file that allows the application to use protected resources in the device.

Each Android api method has a function that involves extracting a resource or modifying something on the device. For an application to access risky methods, it must specify the associated permission in the manifest file. Therefore, we can associate each of these permissions with a set of methods based on the permissions listed in the manifest file.

After collecting all the methods and permissions that could pose a privacy risk, the proposed model assigns them to privacy levels. Each level of privacy will be associated with a specific value representing its inherent privacy risk. Finally, the proposed model applies a formula to calculate the application's overall privacy based on the accumulated privacy levels of all the methods and permissions.

\subsection{System Overview}

We consider three components, each of which plays a crucial role in achieving the final solution. The first component is a mobile application that lets users request privacy analysis for the apps installed on their smartphones. This Android application displays the installed apps and allows users to select the ones they want to analyze. If an application has been previously analyzed, the user can view the privacy score generated by the privacy quantification module and make informed decisions accordingly.

The second component is a server that acts as a gateway between the Android application and the privacy quantification module. Its primary function is to receive http requests made by the Android application on different devices, forward the requests to the privacy quantification module, and return the analysis details and score once the score has been generated.

The third and final component encompasses the privacy quantification module, which undertakes three principal functions. Initially, it receives input from the server and retrieves the application specified by the user. Subsequently, it conducts a comprehensive privacy analysis of the selected application. Finally, it produces a privacy score as part of the evaluation process.

The schematic representation of the proposed infrastructure is illustrated in \autoref{xyz_image}.


\subsection{Mobile Application}
\label{subsec:mobileapp}

The Android application prototype aims to improve the user experience and facilitate easy interaction. The primary function of this app is to communicate with the server and send app analysis requests for the apps installed on the user's device. To store the information retrieved from the server, we have integrated a content provider with the app. The content provider contains only one table with all the necessary details to identify the requested application and the data returned from the server. The data includes the score and pii of the application.

This app has two main activities. The first activity displays a comprehensive list of all the applications installed on the user's device. The list includes both the system apps that come by default with the system, as well as the apps installed by the user. The second activity is shown when the user selects one of the apps from the list. The second activity provides more information about the selected app, such as its score and the piis collected after being analyzed.

Our solution employs a method that does not involve transmission of the actual \ac{apk}s of the requested applications to the server for analysis. Instead, we only send a package name string and a version code integer through a \ac{http} POST request. This approach significantly reduces bandwidth usage. Additionally, we trust only the Google Play Store as the source of the application, not the users who utilize it. This measure ensures that malicious users cannot modify the application for the given package name and version to deceive other users who request the same app. The server database stores the analyzed data, avoiding any unauthorized modifications.

\subsection{Server}
\label{subsec:server}

Our server comprises a Quart\footnote{\url{https://github.com/pallets/quart}} application and a Hypercorn\footnote{\url{https://pgjones.gitlab.io/hypercorn/}} server. We created a server endpoint to handle requests from the mobile application, which accepts an input containing the package name and version code. We used the POST request method to communicate with this endpoint, which is the most suitable option for submitting data to the server for processing or storage. Our aim was to send application data to the server for analysis, facilitating a comprehensive application review. The data can be transmitted in various formats such as Raw data, \ac{xml}, and \ac{json} using a POST request. We opted for \ac{json} due to its simplicity and usability in receiving/sending data, and Python has built-in libraries that can manipulate the values being sent in \ac{json} objects.

Additionally, we use a PostgreSQL database\footnote{\url{https://www.postgresql.org/docs/14/app-initdb.html}} to store the requested applications after analysis to save performance if the same application is requested again. The \ac{json} object expected to be received after a request to our server has two keys or property names, which are \textit{packagename} and \textit{packageversion}. The \textit{packagename} contains the application package name, which usually has the pattern \textit{com.facebook.katana}. The version code that defines a particular version of the application is contained in the \textit{packageversion} key. The version code is a number that is assigned to every \ac{apk} version to keep track of and distinguish between various releases. Higher values denote newer versions, and the version code is usually incremented with each update.

After receiving the values from the user's request, the server executes a database query as its first step. It runs a select query to check if there is an entry that matches the requested package name and version. If there is a match, the server sends the analysis information back to the user's device. However, if there are no entries in the database for a requested app, the server sends a request to the Docker container where the first component of the Privacy Quantification module will download the app.

To receive and execute the components of the Privacy Quantification module, the server uses a Python library called subprocess that simplifies the task of spawning a process, executing a command, and gathering the output.

Once the app is successfully downloaded, the \ac{apk} is saved in the file system, and the Python module sends a confirmation message. If a failure occurs, a notification is generated and included in the user's error response payload. Once the \ac{apk} is present in the file system, and the server confirms the successful download, the second component of the Privacy Quantification module commences execution.

The Application Analysis component obtains the input from the request and initiates the process. If everything goes smoothly, the app is evaluated, and a score ranging from 1 to 100 is sent back to the server from the last component of the Privacy Quantification module, Score Generator. If issues arise during the analysis phase, the server receives the error details and notifies the user.

After preparing the rating for delivery to the user, it is initially stored in the database with the package name, version, and the personally identifiable information detected during the examination. If another user makes an identical request, the server can provide the score without retrieving and scrutinizing the app.

When it finishes storing the data in the server database, a \ac{json} is sent back to the user's device with the privacy score of the app and the extracted sensitive methods and permissions.

\subsection{Privacy Quantification Module}
\label{subsec:privacyquantificationmodule}

The Privacy Quantification module downloads and analyzes the \ac{apk} of a requested application. It then calculates a privacy score based on prepared datasets and extracted privacy-related information, as described previously.

The initial component of our Privacy Quantification module is the Application Download. The first step involves receiving an input from the server that contains the package name and version code of the requested app. The goal of this step is to download the \ac{apk} so that the following steps can analyze it and provide it with a privacy score.

Downloading a mobile application can be a challenging procedure, especially if an automated approach is required. It involves establishing a means of interaction with either the Google Play Store or another third-party application store to facilitate the download of the desired application.

Since the Google Play Store does not have a public official \ac{api} through which users can request information about applications, and the third-party stores can compromise the integrity of the applications, we needed to find a module to communicate with an official store, Google Play Store in this case. After testing available modules, we found one that has had massive recent contributions and usage developed in Python called Google Play python \ac{api}\footnote{\url{https://github.com/Exadra37/googleplay-api}}. Although not developed by Google, it makes requests using Protobufs to an official domain, \textit{android.clients.google.com}, that is used to interact with the Google Play Store. The domain is used to download applications and updates from the Play Store\footnote{\url{https://support.google.com/work/android/answer/10513641?hl=PT&ref_topic=9419964}}.

The Google domain has different endpoints for different kinds of requests to the Google Play Store. For example, the endpoint \textit{/fdfe/details} is used to get details of a specific application inside the Google Play Store. Each function in the module is a request to a specific endpoint.

The flow of this component can be split into six different phases. The first phase is the authentication phase. The authentication is required to make requests to the Google domain by the interface. The interface makes an \ac{http} POST request to the endpoint \textit{auth} with the user's Google account credentials and receives a set of tokens to interact with the domain endpoints.

In the second phase, we get the application details and verify its existence. A GET request is made to the endpoint \textit{/fdfe/details} with the parameters \textit{doc}, which is the package name, and \textit{vc} which is the package version code. From the request, we obtain several pieces of information about the app. \textit{Description}, \textit{offer}, \textit{image details}, \textit{application details}, \textit{rating}, and \textit{related links} are just a few fields we receive in the response. If the app does not exist, we get an error that will be sent to the user's device. Using the field \textit{offer} that we got, we pass to the third phase of the component. The application will use the value of the field to verify if the application is free. If the application is not free but is available, the component generates an error to inform the user that the application specified is not available for analysis.

The fourth phase involves getting the download \ac{url} for the requested application. To download an app, we first need to purchase the app, even if it's free. The purchasing generates a download token that verifies we have legitimate access to that application, and we can download it to our device. The token is required when we call the download function of the interface. To purchase an application, the interface uses an endpoint of the Google domain called \textit{/fdfe/purchase}, which uses an \ac{http} POST request with the package name and version in the parameters. If we get the download token, we go to the next phase; otherwise, we send an error to the user.

In the fifth phase, we call the \textit{/fdfe/delivery} endpoint with the download token to get the download \ac{url}. We use the package name, version code, and download token as parameters in the \ac{http} GET request.

The download \ac{url} obtained from the previous request is responsible for getting the application data and a list of expansion files (usually with a \textit{.obb} extension) that store additional data and resources for an application, typically for a game app. Since these expansion files do not contain any source code of the application, we ignore them and only download the application data.

If any error occurs during the file data acquisition process, we deliver an error message to the user. If everything goes smoothly, we reconstruct the \ac{apk} by using the application data retrieved, and we send a message to the server with the word \textit{Success}. This response means that the server is ready to call the next component, which is the Application Analysis component that will scan the downloaded \ac{apk} file and analyze it in terms of privacy.

%% file: sections/section5.tex
\section{Results}
\label{sec:results}

We conducted a comprehensive evaluation of the most important components in our solution, the Application Analysis, and Score Generator, by using four different applications. For one application, we selected an app that has a high number of permissions and methods that are related to user privacy. On the other hand, we also picked an app with a significantly lower level of privacy and methods, as well as one that was pre-installed on the operating system. Additionally, we created an app prototype with all the dangerous permissions that could potentially compromise the user's privacy. For each of those applications, we will evaluate how the obfuscation affects the final score, compare the app \q{About} page with the final score, and verify if all the methods and permissions extracted for the application reflect the privacy score of it.

\subsection{Shein App}

We chose the Shein app as the subject of our test due to the high number of authorizations it publicly discloses
Shein is a popular shopping app with over 500 million downloads, featuring a wide range of functions that may consume a significant amount of device resources. According to the \q{About} page on the Google Play Store, the Shein app requires access to various features, including the camera for taking photos and videos, precise (GPS) and approximate (network) location, and access to the content on the device.


We analyzed the latest version, 9.7.8, and version code 839. Our analysis revealed that the app contains nine sensitive permissions and 40 methods related to privacy, resulting in a privacy score of 68 out of 100. Some privacy-related methods do not have the required permission in the Android Manifest file. Therefore, our solution gave a weight of 0 to each of these methods since they do not have privileges to access the resource, which does not compromise the user's privacy. We manually calculated the score using the extracted methods and permissions, and obtained a final score of 68, indicating that the Score Generator component is functioning correctly. 

By comparing the extracted methods and permissions with the authorizations shown on the app's \q{About} page, we can observe a correlation between each piece of information that the application is extracting and the methods and permissions used. This means that the final score generated by the methods and permissions extracted is accurate with what is displayed to the user on the Google Play Store.

Moreover, we utilized a tool called \textit{APKID} to verify that the application uses an obfuscator to alter the names of methods and fields
However, our Application Analysis component was able to extract the public methods and generate the score, demonstrating its reliability against obfuscators that can change methods and variable names.


\subsection{Reface App}

In our investigation, we compared Shein with an app named Reface, which is a photo filter application that requires fewer permissions and methods. To verify Reface's permissions, we checked the \q{About} page in the Google Play Store and found that it requires fewer accesses
Our analysis of Reface revealed that it has four permissions and 19 methods related to privacy, which resulted in a final privacy score of 88. However, it should be noted that most methods have a weight of 0 since they do not have the necessary permissions in the Android manifest file.
For instance, Reface lacks permissions such as \textit{ACCESS\_COARSE\_LOCATION} or \textit{ACCESS\_FINE\_LOCATION} in the Android manifest to use the method \textit{getCurrentLocation}. It is essential to understand that this may occur because these methods may belong to a library, and the app is not using them. 


Although we could not identify any obfuscator using the \textit{APKID}, we observed that the classes and variables in the source code have random names, which indicates the use of obfuscation
Despite the obfuscation, the Android classes and their public methods names stay the same, and the Application Analysis component can extract them to calculate the score.


\subsection{Gmail App}

We wanted to test the functionality of our mobile application and see if it can analyze system apps. For this purpose, we decided to test Gmail, which comes pre-installed on all Android devices. We tested version code \textit{63266701} and discovered that despite having more authorizations in the \q{About} page than Shein app
, Gmail had only seven sensitive permissions and 32 methods related to privacy. This means that the authorizations declared in the app \q{About} page can be misleading. We calculated the final privacy score based on the permissions and methods extracted, and it turned out to be 77. However, some methods had a weight of 0 because were not associated with any permissions in the \textit{Androidmanifest.xml} file.


We also ran the \textit{APKID} tool to analyze the application for obfuscators, but it did not identify any obfuscator. Nevertheless, we manually checked and found that the application does have an obfuscation technique that uses letters for the variables and methods names, indicating that an obfuscation technique was used during compile time
. Despite the obfuscation, the application does not obfuscate the Android methods, which allows us to collect them together with the permissions to calculate the final privacy score.


In conclusion, we found that our solution can analyze obfuscated system apps, and we also discovered that the authorizations declared in the app \q{About} page can be misleading. Although Gmail appeared to have more permissions than the Shein app, we found fewer methods and permissions after analysis, resulting in a more private final score.

\subsubsection{Dangerous App}

As part of our research to create a privacy score formula, we created an app called Dangerous App. This app had all sensitive permissions but lacked any privacy-related methods. We then used our formula to calculate the app's privacy score, and it received a final score of 50 out of 100. The result proved that our formula was accurate since permissions contribute to half of the score. However, the app cannot access the resources with only permissions but there is still a risk that the app could add or update code dynamically to add methods and gain access to protected resources.

%% file: sections/section6.tex
\section{Conclusion}
\label{sec:conclusion}

The paper was based on the fact that every app contains all the code and all the functionalities used within the app, and that can be enough to define the information that the app collects about the user. The paper goal was to automatically quantify the privacy of an Android application using the data handled by it. This study has added to the expanding body of knowledge in this field by a thorough analysis of the literature and the creation of an operational framework for assessing privacy in Android applications. 
The framework proposed in this paper uses a combination of prepared datasets of privacy-related API methods and permission and static analysis methodology to analyze the app privacy based on the app source code. The data contained in the datasets was taken from the documentation of the latest version of the Android API. The datasets obtained from this collection were then used, together with the app's source code, to identify the privacy-related methods and permissions used by the app. Once the data handled by the app is identified, we use a formula to generate a privacy score that translates the privacy of the app based on the privacy level of each piece of information collected. Finally, to get the apps from the user's device, we developed an Android application prototype that gathers all the installed applications and sends them to a server that asynchronously handles the requests and analyses them. 
We concluded that we were able to collect the app's methods and permissions that gather the information from the user's device, even if the app is obfuscated. 
We contribute to the existing work by creating updated datasets of privacy-related methods and permissions and establishing an initial formula to calculate the privacy of Android applications.

Although this paper offers a solution to calculate the privacy of Android applications, some of the components can be improved in further research. Due to the automatic requirement of our work, static analysis was the only methodology that fit this situation. However, this type of analysis has limitations. Since it can only use the source code, it does not account for other types of data that are created and handled at runtime, thus limiting the types of data used in the score generation. In addition, obfuscation in the app's source code has a significant impact on how a user can understand what is going on in the code. Due to obfuscation, we have chosen only to gather methods and permissions related to the Android API since they are not typically obfuscated, limiting the information that we can collect from the app source code.

Another improvement for further research that can complement our solution is the analysis of the app privacy policy. The majority of works that use the privacy policy analysis use machine learning. However, due to the ambiguity, the use of contradictions and the non-universal format of the document, it is still difficult to accurately analyze the document automatically. 

Improvements are necessary for the datasets used in the analysis. The accuracy of the collected data by methods and permissions should be more precise. Additionally, privacy levels assigned to the data should be studied and developed to create intermediate privacy levels between the existing ones, resulting in a more precise score.

One of the limitations of our work is the open-source interface that communicates with the Google Play Store. We chose one that better integrates with the other components in our framework. However, it sometimes can produce unexpected errors and may not be the most optimal way to communicate with the Google app store. Both limitations can be easily improved by modifying the source code to our needs.

Finally, another improvement could be in identifying similar piis retrieved by the application. Since the application is not gathering more information when using two methods that collect the same piece of information, we should take that into account and decrease the privacy level of those piis.        